# BASIS OF FINANCIAL ARITHMETIC FROM THE VIEWPOINT OF THE UTILITY THEORY

Krzysztof Piasecki[1]

The main goal of this paper is presentation a modern axiomatic approach to financial arithmetic. At the first, the axiomatic financial arithmetic theory was proposed by Peccati who has introduced the axiomatic definition of the future value. This theory has been extensively developed in past years. Proposed approach to financial arithmetic is based on the financial flow utility concept. This utility function is defined as linear extension of multicriteria comparison determined by the time preference and the capital preference. Then the present value is equal to financial flow utility. Therefore, the law of diminishing marginal wealth utility has been considered as additional feature of the present value. The future value is defined as the inverse of utility function. This definition is a generalization of the Peccati's one. The net present value is given as the unique additive extension of financial flow utility.Moreover, the synergy effect and the diversification effect will be discussed. At the end, the axiomatic present value definition will be specified in three ways.

**Keywords:** *capital synergy effect, diversification, financial arithmetic*, *Gossen's First Law, utility.*

## 1. Introduction

The subject of financial arithmetic is a dynamic estimation of the money value. The fundamental assumption of financial arithmetic is certainty that the money value increases with the time, after which it will be utilized. In general, this assumption is justified by the analysis of the quantity money equation proposed by Irving Fisher [2]. This analysis is based on the additional assumption that the money amount is constant. This is a typical normative assumption. Therefore money value considered in financial arithmetic is called the normative money value. The growth process of normative money value is called the appreciation process of capital. On the other hand, economic-financial practice results generally that the increase in money amount is faster than the increase in production volume. Then we observe a decrease in the real money value. This means that the normative money value cannot be identified with the real money value. This raises question about the essence of the normative value concept. The consequence of this question is another question about the essence of the financial arithmetic basic functions.

---

[1] Poznań University of Economics, Poznan University of Economics,
al. Niepodległości 10, 61-875 Poznań, Poland, k.piasecki@ue.poznan.pl

The answers to this question were sought through the development of an axiomatic theory for financial arithmetic. For any moments set $\Theta \subseteq [0, +\infty[$ Peccati [17] has defined the future value as the function $FV: \Theta \times \mathbb{R} \to \mathbb{R}$ satisfying following properties

$$\forall C \in \mathbb{R}: \quad FV(0, C) = C, \tag{1}$$

$$\forall t_1, t_2 \in \Theta \; \forall C \in \mathbb{R}^+: \; t_1 < t_2 \Rightarrow FV(t_1, C) < FV(t_2, C), \tag{2}$$

$$\forall t \in \Theta \; \forall C_1, C_2 \in \mathbb{R}: \quad FV(t, C_1 + C_2) = FV(t, C_1) + FV(t, C_2). \tag{3}$$

The certainty of money value growth is described above with the axiom (2). The present value has been defined as the function $PV: \Theta \times \mathbb{R} \to \mathbb{R}$ uniquely determined by the identity

$$FV(t, PV(t, C)) = C. \tag{4}$$

Thus was created a coherent theory of financial arithmetic. This approach has been extensively studied, inter alia, in [18; 19]. Current knowledge on the consequences of an axiomatic approach to the future value concept is presented in [10]. On the other hand, the Peccati's theory did not explain the phenomenon of money value growing.

In recent years the concept of financial flow utility has played an important part in the behavioral finance research. This problem is discussed for example by Frederick et al [7], Takahashi [25], Dacey et al [4], Zauberman et al [26], Kim et al [13], Epper et al [6], Killeen [12], Kontek [14], Doyle[5], Piasecki [20; 21; 22] and Han et al [8]. Using this approach we can present the normative value notion in the context of the utility function. This approach sheds new light on the fundamental variables of financial arithmetic.

The main goal of my paper is to explain in terms of utility theory, the basic functions of financial arithmetic: the future value, the present value and the net present value. Obtained in this way definitions will be compared with the axiomatic definitions of present and future values given by Peccati [17]. In order to show the usefulness of the developed theory, we will discuss some properties of the basic financial arithmetic functions. There the law of diminishing marginal wealth utility will be considered as additional feature of the present value. Moreover, the synergy effect and the diversification effect will be discussed.

## 2. Ordered space of financial flows

Let be given the moments set $\Theta \subseteq [0, +\infty[$. In the particular case it may be a capitalization moments set or time non-negative half-line. In the financial market analysis, each payment is represented by the financial instrument described as financial flow $(t, C)$, where the symbol $t \in \Theta$ means flow moment and the symbol $C \in \mathbb{R}$ describes the nominal value of this flow. Each of these financial flows can be executed receivable or matured liabilities. The nominal value of any receivable is non-negative. The debtor's liabilities are always the creditor's receivables. In this situation, each liability value is equal to the minus value of corresponding receivable.

In the first step, we confine our discussion to the set $\Phi^+ = \Theta \times [0, +\infty[$ of all receivables $(t, C)$. Investors determine their preferences on the receivables set. These preferences have some common characteristics.

Reporting on the economic theory basis, von Mises [16] has presented a time-preference rule. This rule says that taking into account the *ceteris paribus* principle, the economic agent will satisfy their needs as quickly as possible. In other words, when the economic agent is faced with two goals characterized by the same subjective value, then it appreciates above this one which can be achieved in less time. In the particular case, this means that the investor, comparing the two payments of equal nominal value, always prefers the payment available quickly. This relation we describe by means of preorder $\succcurlyeq_T$ defined as follows

$$\forall (t_1, C_1), (t_2, C_2) \in \Phi^+ : (t_1, C) \succcurlyeq_T (t_2, C) \Leftrightarrow t_1 \leq t_2. \qquad (5)$$

On the other hand, it is obvious that each economic agent is guided by the rule of capital preference. This rule means that, taking into account the *ceteris paribus* principle, the economic agent will obtain the economic valuables as much as possible. So when economic agent is faced with two economic items available at the same time, he selects the one which is characterized by greater subjective value. In the particular case, this means that the investor, to comparing two simultaneously available payments, always selects the higher payment. This relation we describe by means of preorder $\succcurlyeq_C$ defined as follows

$$\forall (t, C_1), (t, C_2) \in \Phi^+ : (t, C_1) \succcurlyeq_C (t, C_2) \Leftrightarrow C_1 \geq C_2. \qquad (6)$$

Simultaneous taking into consideration both above preorders leads to the final determination of the creditor's preferences $\succcurlyeq$ on the receivables set $\Phi^+$, as multicriteria comparison

$$\forall (t_1, C_1), (t_2, C_2) \in \Phi^+ : (t_1, C_1) \succcurlyeq (t_2, C_2) \Leftrightarrow t_1 \leq t_2 \wedge C_1 \geq C_2. \qquad (7)$$

There exists the utility function $U : \Phi^+ \to [0, +\infty[$ fulfilling the condition

$$\forall (t_1, C_1), (t_2, C_2) \in \Phi^+ : (t_1, C_1) \succcurlyeq (t_2, C_2) \Rightarrow U(t_1, C_1) \geq U(t_2, C_2). \qquad (8)$$

The next step will focused on discussion about the set $\Phi^- = \Theta \times\ ]-\infty, 0]$ of all liabilities. Each debtor's liability $(t, C)$ corresponds to creditor's appropriate receivable $(t, -C)$. Each profit achieved by the creditor is the debtor's expense. This relationship results that debtor's preference defined on set of all the liabilities is the inverse relation to the creditor's preferences defined on a set of appropriate receivables. In this situation, the debtor's preferences $\succcurlyeq$ on the set $\Phi^-$ of all liabilities are defined by equivalency

$$\forall (t_1, C_1), (t_2, C_2) \in \Phi^- : (t_1, C_1) \succcurlyeq (t_2, C_2) \Leftrightarrow (t_2, -C_2) \succcurlyeq (t_1, -C_1). \tag{9}$$

The comparison of equation (7) and (9) leads to the final determination of the debtor's preferences $\succcurlyeq$ on the set $\Phi^-$ of all liabilities, as as multicriteria comparison

$$\forall (t_1, C_1), (t_2, C_2) \in \Phi^- : (t_1, C_1) \succcurlyeq (t_2, C_2) \Leftrightarrow t_1 \geq t_2 \wedge C_1 \geq C_2. \tag{10}$$

There exists the utility function defined on the set of all liabilities. From a financial point of view, each receivable is more useful than any liability. Thus we can say that

$$\forall \big((t_1, C_1), (t_2, C_2)\big) \in \Phi^+ \times \Phi^- : U(t_1, C_1) \geq U(t_2, C_2). \tag{11}$$

Fulfillment of this condition can be obtained through the assumption that the utility of any liability is non-positive[2]. Therefore we can say that there exists the utility function $U: \Phi^- \to\ ]-\infty, 0]$ fulfilling the condition

$$\forall (t_1, C_1), (t_2, C_2) \in \Phi^- : (t_1, C_1) \succcurlyeq (t_2, C_2) \Rightarrow U(t_1, C_1) \geq U(t_2, C_2). \tag{12}$$

To sum up the previous discussion we can conclude, that financial preferences $\succcurlyeq$ are determined on the set $\Phi = [0, +\infty[\ \times \mathbb{R} \to \mathbb{R}$ of all financial flows. This relation is determined by alternative of multicriteria comparisons in the following way

$$\forall (t_1, C_1), (t_2, C_2) \in \Phi : (t_1, C_1) \succcurlyeq (t_2, C_2) \Leftrightarrow$$

$$\Leftrightarrow (t_1 \geq t_2 \wedge 0 \geq C_1 \geq C_2) \vee (t_1 \leq t_2 \wedge C_1 \geq C_2 \geq 0) \vee (C_1 \geq 0 \geq C_2) \tag{13}$$

This preorder is not linear. There exists the utility function $U: \Phi \to \mathbb{R}$ fulfilling the condition

$$\forall (t_1, C_1), (t_2, C_2) \in \Phi : (t_1, C_1) \succcurlyeq (t_2, C_2) \Rightarrow U(t_1, C_1) \geq U(t_2, C_2). \tag{14}$$

About this function we will also assume that its values are determined by receivables utility function (8) or by liabilities utility function (12). Defined in this way, the utility function can be subjective [4]. This indicates the possibility that financial preferences model introduced above may be applied for behavioural finance.

---
[2] Negative utility notion was discussed in the papers [1], [3] and [23].

### 3. Financial flow utility

Let us now examine the properties of utility functions determined above. Comparison of domains and codomains of utility functions described in (8) and (12) implies that

$$\forall t \in \Theta: \quad U(t, 0) = 0 \ . \tag{15}$$

The preorder $\succcurlyeq$ determines the strict order $\succ$ defined on the set $\Phi$ by the equivalence

$$\forall (t_1, C_1), (t_2, C_2) \in \Phi: (t_1, C_1) \succ (t_2, C_2) \Leftrightarrow$$

$$\Leftrightarrow \left( \left( (t_1, C_1) \succcurlyeq (t_2, C_2) \right) \wedge \sim \left( (t_2, C_2) \succcurlyeq (t_1, C_1) \right) \right). \tag{16}$$

This strict order is determined by the alternative of multicriteria comparisons

$$\forall (t_1, C_1), (t_2, C_2) \in \Phi: (t_1, C_1) \succ (t_2, C_2) \Leftrightarrow$$

$$\Leftrightarrow (t_1 \geq t_2 \wedge 0 \geq C_1 > C_2) \vee (t_1 > t_2 \wedge 0 \geq C_1 \geq C_2) \vee$$

$$\vee (t_1 \leq t_2 \wedge C_1 > C_2 \geq 0) \vee (t_1 < t_2 \wedge C_1 \geq C_2 \geq 0). \tag{17}$$

On the other hand, we have here

$$\forall (t_1, C_1), (t_2, C_2) \in \Phi: (t_1, C_1) \succ (t_2, C_2) \Rightarrow U(t_1, C_1) > U(t_2, C_2). \tag{18}$$

Comparing (16), (17) and (18) we find that the utility is an increasing function of the flow nominal value. Thus we can say that

$$\forall t \in \Theta \, \forall C_1, C_2 \in \mathbb{R}: \ C_1 > C_2 \Rightarrow U(t, C_1) > U(t, C_2) \ . \tag{19}$$

Thus, for any fixed moment $t \in \Theta$ we can determine the inverse function $U_0^{-1}(t, \cdot): \mathbb{R} \to \mathbb{R}$. In addition, comparing (17) and (18) we get

$$\forall C > 0 \, \forall t_1, t_2 \in \Theta: t_1 < t_2 \Rightarrow U(t_1, C) > U(t_2, C) \ . \tag{20}$$

$$\forall C < 0 \, \forall t_1, t_2 \in \Theta: t_1 < t_2 \Rightarrow U(t_1, C) < U(t_2, C) \ . \tag{21}$$

The conventional issue is the calibration of the utility function value. Here we assume that the utility of immediate financial flow is equal to the nominal value of this flow. This assumption is written as a boundary condition

$$\forall C \in \mathbb{R}: \ U(0, C) = C \ . \tag{22}$$

All these utility function properties will be used to study the properties of the financial arithmetic basic models.

## 4. Future and present values

For the preorder ⩾ defined by the equivalence (13) we determine its linear closure ⊒ in following way

$$\forall (t_1, C_1), (t_2, C_2) \in \Phi: (t_1, C_1) \sqsupseteq (t_2, C_2) \Leftrightarrow U(t_1, C_1) \geq U(t_2, C_2). \quad (23)$$

This preorder appoints the relation ≡ of financial flow equivalence. We have here

$$\forall (t_1, C_1), (t_2, C_2) \in \Phi: (t_1, C_1) \equiv (t_2, C_2) \Leftrightarrow U(t_1, C_1) = U(t_2, C_2). \quad (24)$$

If two financial flows are equally useful then we consider them as equivalent. If two financial flows are equivalent then first one is called the equivalent of the second one. Each nominal value of any financial flow equivalent is identified as a normative value of this flow.

Analysis of conditions (19) and (20) leads us to formulate the appreciation principle. This principle states that the receivable normative value increases with time, after which this receivable will be paid. In this way, the utility theory confirms the usefulness of fundamental financial arithmetic axiom stating that the value of money value increases with time.

The above-described concept of normative value can be included in the framework of formal model. Let there be given an immediate financial flow with the nominal value $C \in \mathbb{R}$. This financial flow is uniquely assigned by the pair $(0, C)$. At any time $t \in \Theta$ the normative value of discussed flow is equal to $C_t$. In accordance with the definition (24) of financial flow equivalence and boundary condition (22) we have here the identity

$$C = U(0, C) = U(t, C_t). \quad (25)$$

Under the condition (19), for a fixed moment $t \in \Theta$ we uniquely determine the normative value

$$C_t = FV(t, C) = U_0^{-1}(t, C). \quad (26)$$

Defined in this way function $FV: \Theta \times \mathbb{R} \to \mathbb{R}$ is called the future value. In the general case, this function has the following properties: (1), (2) and

$$\forall t \in \Theta: \ FV(t, 0) = 0, \quad (27)$$

$$\forall (t_1, C), (t_2, C) \in \Phi^-: \ t_1 < t_2 \Rightarrow FV(t_1, C) > FV(t_2, C), \quad (28),$$

$$\forall (t, C_1), (t, C_2) \in \Phi: C_1 < C_2 \Rightarrow FV(t, C_1) < FV(t, C_2). \qquad (29),$$

It is easy to check that condition (29) is a generalization of condition (3). This means that, the future value function defined in this chapter is a generalization of the future value in perspective proposed by Peccati [17]. Future value function can be represented by the identity

$$FV(t, C) = C \cdot s(t, C), \qquad (30)$$

where the function $s: \Phi \to [0, +\infty[$ is called appreciation factor. The appreciation factor is increasing time function fulfilling boundary condition:

$$s(0, C) = 1. \qquad (31)$$

Appreciation factor describes the process of capital relative appreciation. If this factor is an increasing function of positive capital value is then we have to deal with the capital synergy effect. Capital synergy effect consists in that the increase in positive capital value causes the increase in the relative speed of the appreciation.

Another object of our investigation will be any financial flow $(t, C)$. For this flow we can determine its equivalent $(0, C_0)$. The nominal value $C_0$ of this equivalent is called the present value and it is denoted by $PV(t, C)$. In accordance with the definition (24) of financial flow equivalence and boundary condition (22) we have here the identity

$$C_0 = PV(t, C) = U(0, C_0) = U(t, C). \qquad (32)$$

Present value of any financial flow is equal to its utility. This statement fully explains the essence of the present value concept. On the other hand, this present value interpretation is not without some formal problems, which will be discussed later. Now let's focus our attention to the formal properties of function $PV: \Phi \to \mathbb{R}$ defined by the identity (32). We have here

$$\forall C \in \mathbb{R}: \quad PV(0, C) = C, \qquad (33)$$

$$\forall t \in \Theta: \quad PV(t, 0) = 0, \qquad (34)$$

$$\forall (t_1, C), (t_2, C) \in \Phi^+: \quad t_1 < t_2 \Rightarrow PV(t_1, C) > PV(t_2, C), \qquad (35)$$

$$\forall (t_1, C), (t_2, C) \in \Phi^-: \quad t_1 < t_2 \Rightarrow PV(t_1, C) < PV(t_2, C), \qquad (36)$$

$$\forall (t, C_1), (t, C_2) \in \Phi: C_1 < C_2 \Rightarrow PV(t, C_1) < PV(t, C_2). \qquad (37)$$

Present value function can be represented by the identity

$$PV(t, C) = C \cdot v(t, C), \qquad (38)$$

where discounting factor $v: \Phi \to \,]0;1]$ s decreasing time function fulfilling boundary condition:

$$v(0, C) = 1. \qquad (39)$$

Comparison of conditions (26) and (32) shows that defined in this section the future value and the present value satisfy the condition (4). This means that the present value function defined here is a generalization of the present value present value in in perspective proposed by Peccati [17].

Let consider again the pair of equivalent financial flows $(0, C_0)$ i $(t, C_t)$. From equations (26), (30), (32) and (38) we have here

$$C_t = FV(t, C_0) = C_0 \cdot s(t, C_0), \qquad (40)$$

$$C_0 = PV(t, C_t) = C_t \cdot v(t, C_t). \qquad (41)$$

Comparison of the last two equations leads to the relation

$$s(t, C_0) = \frac{1}{v(t,C_t)} = \frac{1}{v(t,FV(t,C_0))}.$$

In this way, we have shown that the appreciation and discounting factors determined by the same utility function satisfy the condition

$$s(t, C) = \frac{1}{v(t,FV(t,C))} = \frac{1}{v(t,C \cdot s(t,C))}. \qquad (42)$$

## 5. Net present value

Let be given the set $\Phi = \Theta \times \mathbb{R}$ of all financial flows. Each sequence of financial flows is called financial investment. Then any investment $\check{X}$ we describe as the multiset [9] of financial flow:

$$\check{X} = \{(t_j, C_j) \in \Phi : j = 1, 2, \dots, n, \dots\}. \qquad (43)$$

It means that in any investment may have different financial flows with identical moments and identical nominal value. Each investment containing exactly one financial flow is called the simple investment. The family of all investment we denote by the symbol $\mathbb{F}$.

The combination of investments pair is the investment consisting all financial flows of both investments. In accordance with above, the combination $\check{X} \uplus \check{Y}$ of investments pair $(\check{X}, \check{Y})$ is given as the multiset sum [24]:

$$\check{X} \uplus \check{Y} = \{(t,C): (t,C) \in \check{X} \lor (t,C) \in \check{Y}\}. \tag{44}$$

In the family $\mathbb{F}$ of all investments we distinguish the subfamily $\mathbb{F}_1$ of all simple investments. Any investment can be presented as a countable multiset sum of simple investments. On the other hand the family $\mathbb{F}_1$ of all simple investment and the set $\Phi$ of all financial flows are isomorphic. Thus the preorder $\sqsupseteq$ on the set $\Phi$ of all financial flows determines the linear preorder $\gtrsim$ on the set $\mathbb{F}_1$ of all simple investments. This preorder is defined by the equivalence

$$\forall \{(t_1,C_1)\}, \{(t_2,C_2)\} \in \mathbb{F}_1: \{(t_1,C_1)\} \gtrsim \{(t_2,C_2)\} \Leftrightarrow (t_1,C_1) \sqsupseteq (t_2,C_2). \tag{45}$$

It means that there exists the utility function $V: \mathbb{F}_1 \to \mathbb{R}$ determined by the identity

$$V(\{(t,C)\}) = U(t,C). \tag{46}$$

Let us take into account any extension of the utility function $V: \mathbb{F} \to \mathbb{R}$. This extension we determine with the postulate that the utility function is additive [11]. This postulate is consistent with the finance practice, where the value of capital is calculated as the sum of its components value. It follows that any investment utility function $V: \mathbb{F} \to \mathbb{R}$ should satisfy the following additivity condition

$$\forall \check{X}, \check{Y} \in \mathbb{F}: \quad V(\check{X} \uplus \check{Y}) = V(\check{X}) + (\check{Y}). \tag{47}$$

Conditions (32), (45) and (46) are sufficient that the utility of investment $\check{X} \in \mathbb{F}$ is uniquely designated as the net present value. We have here

$$\forall \check{X} \in \mathbb{F}: \quad \text{NPV}(\check{X}) = V(\check{X}) = \sum_{(t,C) \in \check{X}} \text{PV}(t,C). \tag{48}$$

Using the above determined utility function $NPV: \mathbb{F} \to \mathbb{R}$ we can extend the linear preorder $\gtrsim$ to the family $\mathbb{F}$ of all investments. Here we have

$$\forall \check{X}, \check{Y} \in \mathbb{F}: \quad \check{X} \gtrsim \check{Y} \Leftrightarrow NPV(\check{X}) \geq NPV(\check{Y}). \tag{49}$$

This preorder is also linear one. This is a natural preorder applied in finance practice. This preorder sets the equivalence relation $\equiv$ on the family $\mathbb{F}$ of all investments. We have here

$$\forall \check{X}, \check{Y} \in \Lambda: \quad \check{X} \equiv \check{Y} \Leftrightarrow NPV(\check{X}) = NPV(\check{Y}). \tag{50}$$

This relation is a generalization of the financial flow equivalence defined by the condition (24). In this way in the last two chapters, the functions of the present value, future value and net present value have been defined on the basis of the general utility theory. This fact will be used in discussions about the specific properties of these functions

## 6. Gossen's First Law

The Gossen's First Law says that the marginal wealth utility is diminishing [2]. Now let us examine the consequence of accepting the assumption that the utility function $U: \Phi^+ \rightarrow [0, +\infty[$ defined by (8) fulfils the law of diminishing marginal wealth utility. Then for any present value function we can write

$\forall (t, C_1), (t, C_2) \in \Phi^+ \; \forall \alpha \in \;]0; 1[\;:$

$$\alpha \cdot PV(t, C_1) + (1 - \alpha) \cdot PV(t, C_2) < PV(t, \alpha \cdot C_1 + (1 - \alpha) \cdot C_2). \tag{51}$$

Present value is a concave function over the set of all positive capital values. This allows proving the following theorem.

Theorem 1: The fulfillment of the Gossen's First Law is necessary and sufficient condition for this that the future value $FV: \Phi \rightarrow [0, +\infty[$ variability reveals the capital synergy effect.

Proof: Let be given fixed moment $t \in \Theta$. In the inequality (51) we substitute $C_2 = 0$. For any positive value $C_3 < C_1$ we have here

$$\frac{C_3}{C_1} \cdot PV(t, C_1) + \left(1 - \frac{C_3}{C_1}\right) \cdot PV(t, 0) < PV\left(t, \frac{C_3}{C_1} \cdot C_1 + \left(1 - \frac{C_3}{C_1}\right) \cdot 0\right),$$

which together with (37) leads to

$$v(t, C_1) = \frac{PV(t, C_1)}{C_1} < \frac{PV(t, C_3)}{C_3} = v(t, C_3). \tag{52}$$

Thus discounting factor is decreasing function of positive nominal value. On the other hand, if discounting factor fulfils (52) then in accordance with (38), functions $PV(t, C)$ and $f(C) = C - PV(t, C)$ are increasing functions of positive nominal value. It implies that the condition (51) is fulfilled. The inequality (51) is necessary and sufficient condition for the inequality (52). Comparison of equation (42) and (29) shows it that discounting factor decreases iff appreciation factor increases. This conclusion ends the proof.□

## 7. Investments diversification

The investment composed from financial flows with an identical flow moment is called a portfolio. Then, the net total nominal value of those financial flows is called the portfolio future value. The common flow moment is called the portfolio maturity.

The investments diversification is understood as the resources allocation among different investments. The diversification principle has popularized with the portfolio theory introduced by Markowitz [15]. This principle states that diversification should be preferred. The formal model of the diversification principle is the condition

$$\{(t, C_1), (t, C_2)\} \succsim \{(t, C_1 + C_2)\}. \tag{53}$$

At the beginning we examine the impact of accepting the diversification principle on the present value. In accordance with (48) and (49), the condition (53) is equivalent to the inequality

$$PV(t, C_1) + PV(t, C_2) \geq PV(t, C_1 + C_2). \tag{54}$$

If the present value fulfils above inequality then the conditions (15) and (32) imply

$$PV(t, C) + PV(t, -C) \geq PV(t, C - C) = 0. \tag{55}$$

Then for $C > 0$ we have

$$C \cdot v(t, C) - C \cdot v(t, -C) \geq 0,$$

which allows us to write

$$v(t, C) \geq v(t, -C). \tag{56}$$

It shows that under the diversification principle influence liabilities are discounted more strongly than receivables.

Now we examine the impact of accepting the diversification principle on the future value.

Theorem 2: If the condition (54) is fulfilled then the future value $FV: \Phi \to [0, +\infty[$ satisfies the inequality

$$FV(t, C_1) + FV(t, C_2) \leq FV(t, C_1 + C_2). \tag{57}$$

Proof: For $(t, C_1), (t, C_2) \in \Phi$ using the condition (26), (32) and (54) we obtain

$$PV\big(t, FV(t, C_1 + C_2)\big) = U\big(t, U_0^{-1}(t, C_1 + C_2)\big) = C_1 + C_2 =$$

$$= PV(t, FV(t, C_1)) + PV(t, FV(t, C_2)) \geq PV(t, FV(t, C_1) + FV(t, C_2)),$$

which together with inequality (37) gives (57). □

This result indicates that the capital growth rate increases with the increase of its value. Moreover, let us note that for $C_2 < 0$ the condition (57) describes financial leverage effect.

Using the equations (15) and (26) we can say

$$FV(t, C) + FV(t, -C) \leq FV(t, C - C) = 0. \tag{58}$$

Then for $C > 0$ we have

$$C \cdot s(t, C) - C \cdot s(t, -C) \leq 0,$$

which implies

$$s(t, C) \leq s(t, -C). \tag{59}$$

This result indicates that the receivables growth rate is less or equal to the liabilities growth rate. This is fully consistent with the interpretation of inequality (56).

## 8. Diversification neutrality

In the particular case, assessing future financial flows, we can to ignore the potential benefits achieved through investment diversification. This is known as the diversification neutrality. The formal model of the diversification neutrality is the condition

$$\{(t, C_1), (t, C_2)\} \equiv \{(t, C_1 + C_2)\}. \tag{60}$$

In accordance with the equations (48) and (50), the condition (60) jest equivalent to the identity

$$PV(t, C_1) + PV(t, C_2) = PV(t, C_1 + C_2). \tag{61}$$

Let us consider any present value $PV: \Phi \to \mathbb{R}$ fulfilling the conditions (33) and (35). Then in [18], [19] it was shown that the condition (61) is necessary and sufficient for it that the present value is given by the identity

$$PV(t, C) = C \cdot v(t), \tag{62}$$

where discounting factor $v: \Theta \to \,]0; 1]$ jest nonincreasing time function satisfying

$$v(0) = 1. \tag{63}$$

The present value determined in this way is a linear capital function. It means that diversification neutrality is necessary and sufficient condition for rejection of the Gossen's First Law. Such rejection only means that in assessing the financial flows we ignore the diminishing marginal utility effect.

The condition (42) implies that the present value $PV: \Phi \to \mathbb{R}$ fulfills the conditions (62) and (63) iff when the future value $FV: \Phi \to \mathbb{R}$ is given by

$$FV(t, C) = C \cdot s(t), \qquad (64)$$

where the appreciation factor $s: \Phi \to [0, +\infty[$ is increasing time function fulfilling boundary condition:

$$s(0) = 1. \qquad (65)$$

It means that diversification neutrality is necessary and sufficient condition for rejections of the capital synergy effect. Such rejection only means that in assessing the financial flows we ignore the capital synergy effect.

The condition (64) is equivalent to the condition (3). Thus the condition (3) is equivalent to condition (60) of diversification neutrality. All this means that in the Peccati's definition of future value the diversification neutrality is assumed implicitly.

## 8. Definition of generalized present value

In this work it was shown that application of „classical" financial arithmetic rules means that the effects of the diminishing marginal utility, the capital synergy and the diversification principle are omitted in financial flow assessing. This observation encourages to the search for useful generalizations of future value definition.

The generalization of the Peccati's future value definition can be achieved by replacing the condition (3) by more general inequality (29). Moreover, inequality (29) is a necessary condition for retaining the diversification principle. Thus generalized future value is defined as any function $FV: \Phi \to \mathbb{R}$ fulfilling the condition (1), (2) and (29).

On the other hand, the Gossen's First Law is expressed by means of the present value function. Thus it is convenient to take the present value definition as a formal basis for financial arithmetic. Then generalized future value definition should be replaced by equivalent definition of generalized present value which is defined as any function $PV: \Phi \to \mathbb{R}$ fulfilling the condition (33), (35) and (37).

The inequality (37) is a necessary condition for retaining the diversification principle. There was show that the diversification principle is the sufficient condition for taking into account the synergy effect. On the other hand the synergy effect is equivalent to law (51) of diminishing marginal utility. These observations lead to undertake detailed studies of three variants of the generalized present value definition:

- generalized present value $PV: \Phi \to \mathbb{R}$ given as any function fulfilling the conditions (33), (35) and (37),
- generalized present value $PV: \Phi \to \mathbb{R}$ given as any function fulfilling the conditions (33), (35) and (54),
- generalized present value $PV: \Phi \to \mathbb{R}$ given as any function fulfilling the conditions (33), (35) and (51).

Such may be the topics of future studies of prescriptive models for financial arithmetic.

## 9.Conclusions

Presented above the relationship between the wealth utility and the present value shows the logical consistency of formal economics and finance models. Finding these similarities is especially important now when we have become participants in the global financial crisis triggered of financial management in isolation from the fundamental bases created by the economy.

The present value notion is subjective in nature because of it is identical with financial flow utility. In this situation, we obtain a theoretical foundation for the construction of behavioral finance models using subjective evaluation for determining the present value.

Let us notice that the financial arithmetic field is increasingly goes beyond the interest theory domain. In passing this work is worth to notice that the area of financial arithmetic is increasingly beyond the domain of theory of interest. In this situation, the financial arithmetic should be treated as a subjective extension of the interest theory which is based on objective premises. In this paper was show that this extension is important.